\documentclass[letterpaper, 10pt, conference]{ieeeconf} 

\IEEEoverridecommandlockouts                              
\usepackage{amsmath}
\usepackage[final]{graphicx}
\usepackage{epsfig}
\usepackage{subfigure}
\usepackage{amsfonts}
\usepackage{latexsym,amssymb}
\usepackage[center]{caption}
\usepackage{color}
\usepackage{url}
\usepackage[belowskip=-5pt,aboveskip=0pt]{caption}
\usepackage[T1]{fontenc}
\usepackage{pslatex}

\newcommand{\ignore}[1]{}

\newtheorem{alg}{Algorithm}
\newtheorem{prop}{Proposition}

\title{\LARGE \bf
On the Efficiency-vs-Security Tradeoff in the Smart Grid
}

\author{Yara Abdallah, Zizhan Zheng, Ness B. Shroff and Hesham El Gamal
\thanks{Yara Abdallah, Zizhan Zheng, Ness B. Shroff and Hesham El Gamal are with the Department of Electrical and Computer Engineering,
        The Ohio State University, Columbus, Ohio, USA.
        {\tt\small \{abdallay, zhengz, shroff, helgamal\}@ece.osu.edu}}%
\thanks{This work has been funded in part by the Army Research Office MURI award W911NF-08-1-0238 and National Science Foundation awards ECCS-1232118.}%
}

\begin{document}

\maketitle
\thispagestyle{empty}
\pagestyle{empty}

\begin{abstract}
The smart grid is envisioned to significantly enhance the efficiency of energy consumption, by utilizing two-way communication channels between consumers and operators. For example, operators can opportunistically leverage the delay tolerance of energy demands in order to balance the energy load over time, and hence, reduce the total operational cost. This opportunity, however, comes with security threats, as the grid becomes more vulnerable to cyber-attacks. In this paper, we study the impact of such malicious cyber-attacks on the energy efficiency of the grid in a simplified setup. More precisely, we consider a simple model where the energy demands of the smart grid consumers are intercepted and altered by an active attacker before they arrive at the operator, who is equipped with limited intrusion detection capabilities. We formulate the resulting optimization problems faced by the operator and the attacker and propose several scheduling and attack strategies for both parties. Interestingly, our results show that, as opposed to facilitating cost reduction in the smart grid, increasing the delay tolerance of the energy demands potentially allows the attacker to force increased costs on the system. This highlights the need for carefully constructed and robust intrusion detection mechanisms at the operator.
\end{abstract}

\section{Introduction}\label{sec:introduction}

Over the past few years, the smart grid has received considerable momentum, exemplified in several regulatory and policy initiatives, and research efforts (see for example \cite{Moslehi2010a,Lui2010} and the references therein). Such efforts have addressed a wide range of topics spanning energy generation, transportation and storage technologies, sensing, control and prediction, and cyber-security~\cite{McDaniel2009}.

Demand response/load balancing and energy storage are two promising directions for enhancing energy efficiency in the smart grid. Non-emergency demand response has the potential of lowering real-time electricity prices and reducing the need for additional energy sources. The basic idea is that, by utilizing two-way communication channels, the \emph{emergency level} of each energy demand (at the end-users or central distribution stations) is sent to the grid operator that, in turn, \emph{schedules} these demands in a way that \emph{flattens} the load\ignore{ assuming sufficient energy storage capabilities at the consumers}. This potential gain, however, comes at the expense of the security threat posed by the vulnerability of the communication channels to interception and impersonation.

\begin{figure}
 \centering
 \includegraphics[width=0.45\textwidth]{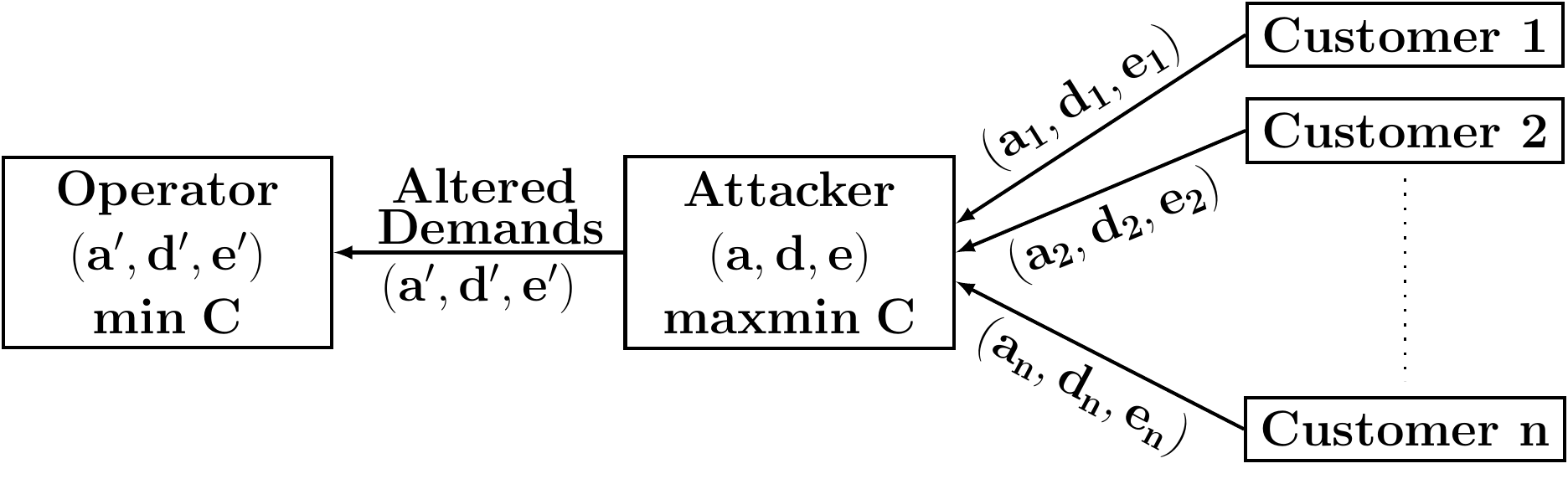}
 \caption{A system model for a smart grid in the presence of a single attacker. The forward channels between the consumers and the grid operator are fully compromised by the attacker. $(a,d,e)$ is the vector of the start times, deadlines and energy requirements of the consumer demands, respectively.}
 \label{fig:system_model}
\end{figure}

This paper is, to the best of our knowledge, the first attempt to characterize the impact of cyber-attacks on the smart grid, in terms of its energy efficiency. More specifically, we propose a novel model that captures the above scenario in the presence of a single attacker. Our model of the smart grid, similar to \cite{Koutsopoulos2010}, includes a grid operator and $n$ consumers that are capable of energy storage, harnessing the potential cost savings in the smart grid. Each consumer has a \emph{single} energy demand that includes the amount of energy the consumer requests, the service start time, and the \emph{deadline} by which the requested energy should be delivered. The consumers send their demands, simultaneously, over separate communication channels to the operator. The grid operator attempts to schedule these demands so as to balance the load across a finite period of time, and hence \emph{minimize} the total cost paid to serve these demands. In our model, we also assume the presence of a single attacker who is fully capable of intercepting and altering the consumer demands before they arrive at the operator, as shown in Figure \ref{fig:system_model}. The end goal of the attacker, as opposed to the operator, is to \emph{maximize} the operational cost paid by the system for these demands, hence reducing the energy efficiency of the system. We differentiate between two scenarios. The first corresponds to a naive operator who fully trusts the incoming energy demands, whereas in the second, a simple intrusion detection mechanism (that will be discussed later) is assumed to be deployed by the operator. The attacker's desire to remain undetected imposes more limitations on its capabilities, and hence, reduces the potential harm. This desire can be justified, for example, by considering the long-term performance of the grid, i.e., successive instances of the problem considered in our model, where in each instance $n$ energy demands are issued and altered by an attack. From this perspective, one can envision scenarios where the total impact of successive attacks is more damaging when the attacker remains undetected.

Based on the aforementioned assumptions, we first formulate the optimization problems faced by the operator and the attacker. For the operator, when being oblivious to any attacks, a minimization problem needs to be solved. On the other hand, the attacker is aware of the optimal strategy employed by the operator, and hence, a maximin optimization problem needs to be solved. In our formulation, we limit the attack's strength by the number of energy demands he is capable of altering, \emph{without} being detected. For the case where the attacker is capable of altering \emph{all} of the consumer's energy demands (the attacks thus reach their full potential and force the system to operate at the maximum achievable total cost), we show that the attacker's maximin problem is reduced to a maximization problem. Our main contribution can be summarized as follows.

\begin{itemize}
  \item For both the operator and an unlimited attacker, we propose optimal offline strategies (Section \ref{sec:optimal_strategies}). The gap between the two indicates the maximum damage that an attack possibly causes. We also provide efficient online strategies for both of them, which are more practical in terms of operability and indicate a lower bound on the possible damage due to an unlimited attack.
  \item For more limited attacks (Section \ref{sec:limited_attacks}), we use a simple greedy algorithm to arrive at a lower bound on the resulting total cost in terms of the flexibility allowed to the stealthy attacker for altering the demands. Additionally, we provide a Dynamic Programming-based algorithm that computes an upper bound on the total cost achieved by such attacks.
  \item We provide numerical results that support our theoretical claims under different scenarios (Section \ref{sec:numerical}). In these studies, we compare the average system performance in the presence/absence of attacks with the expected system performance when the delay tolerance of the jobs is not exploited by the operator (resembling the current electric gird where the communication infrastructure is absent). Moreover, we show the trade-off between the strength of the intrusion detection at the operator and the reduction in the system's efficiency due to stealthy attacks.
  \item From our analysis and numerical results, we conclude that {\bf in the absence of security threats} an increase in the delay tolerance of the energy demands increases the energy efficiency of the system, as expected, since the smart grid's operator is offered more scheduling opportunities. On the other hand, with a limited defense mechanism at the operator, this increase offers a similar opportunity to the attacker to force costs {\bf even higher than those incurred by the regular grid}, transposing the purpose of the communication capabilities provided to the consumers.
\end{itemize}

\section{Problem Formulation}\label{sec:formulation}

In this paper, we consider the control and optimization framework first proposed in~\cite{Koutsopoulos2010} for the demand side of the smart grid. This framework assumes a central controller and $n$ energy consumers that send their energy service demands to the controller using perfect channels. We consider a time-slotted system with this model and add to it a single active attacker, that is capable of intercepting and altering the consumer demands. Let $J=\{1,\ldots,n\}$ denote the set of energy demands. The $j^{th}$ energy demand is composed of the tuple $(a_j,d_j,e_j)$, where $a_j,d_j \in \mathbb{N}^+$ denote the the demand's arrival time and deadline, respectively, $e_j \in \mathbb{R}^+$ denotes the requested total energy by the consumer, and $a_1 \leq a_2 \leq \ldots \leq a_n$. Each energy demand is sent to the controller over a perfect channel that is fully intercepted by the attacker. Hence the attacker can substitute each demand $(a_j,d_j,e_j)$ by $(a'_j,d'_j,e'_j)$, which are then received by the controller. For ease of notation, we define $a = [a_1,\ldots,a_n]$. $a',d,d',e,e'$ are defined similarly.

Upon receiving the $n$ (altered) demands, an \emph{admissible} schedule of these jobs is to be determined by the controller. A schedule is admissible if each job is served its requested energy upon or after its arrival and before or upon its deadline (job preemption is allowed). Letting $T = \max_j d'_j$, a schedule is given by $S \in \mathbb{R^+}^{n \times T}$, where $s_{jt}$ denotes the amount of energy allocated to job $j$ in time slot $t$. Let $E_S(t)$ be the total energy consumed at time slot $t \in [0,T]$ under the schedule $S$, i.e., $E_S(t) = \sum_{j\in J} s_{jt}$. Let $C(E_S(t))$ denote the cost paid for the total energy consumed at time slot $t$ with schedule $S$, where $C: \mathbb{R}^+ \rightarrow \mathbb{R}^+$ is assumed to be non-decreasing and convex. The convexity assumption implies that, as the demand increases, the differential cost at the operator increases, i.e., serving each additional unit of energy to increasing demand becomes more expensive~\cite{Koutsopoulos2010}. Accordingly, the controller attempts to find an admissible schedule that \emph{balances} the load over $[0,T]$. The optimization problem at the controller side is then defined as follows:
\vspace{-0.05in}
\begin{align}
C_{min}(a',d',e')  &=  \min_{S}
 \sum_{t=1}^{T} C(E_S(t)) \notag\\
 \text{s.t.}\qquad
  &s_{jt} \geq 0, \ \forall j \in J, \forall t \in [0,T],\notag\\
  &\sum_{t=1}^T s_{jt} = e'_j, \ \forall j \in J,\notag\\
  &s_{jt} = 0, \ \forall t < a'_j, t > d'_j,  \forall j \in J.\tag{Pmin}\label{pr:min}
\end{align}

On the other hand, the attacker attempts to find appropriate values of $a',d',e'$ such that the cost achieved by the legitimate controller is maximized, \emph{without} being detected (see Figure \ref{fig:system_model}). The intrusion detection capability at the controller is modeled as the number of energy demands the attacker is capable of altering without being detected. This threshold is known a priori to all parties and the attacker solves:
\vspace{-0.05in}
\begin{align}
C_{maxmin}(a,d,e,\beta) & = \max_{a',d',e',J^*}
 C_{min}(a',d',e') \notag\\
 \text{s.t.} \hspace{3em}\qquad
&    a'_j, d'_j \in \mathbb{N}^+, \hspace{2em} \forall j \in J,\notag\\
&    e'_j = e_j, \hspace{2em} \forall j \in J,\notag\\
&    a'_j \geq a_j, d'_j \leq d_j, \hspace{2em} \forall j \in J,\notag\\
&    |J^*| \leq \beta n, \tag{Pmaxmin}\label{pr:maxmin}
\end{align}%
where $\beta \in \mathbb{R}, 0 \leq \beta \leq 1, \beta n \in \mathbb{N}^+$ and \begin{equation}\label{eq:maximin_constraint}
    J^* = \{j \in J \colon a'_j \neq a_j \text{ or } d'_j \neq d_j\}.
\end{equation}

In the above formulation, $J^*$ denotes the set of jobs altered by the attacker, and $\beta$ denotes the {\it fraction} of jobs that can be altered without being detected. The remainder of the constraints imply that, if the energy requirement of a job is not satisfied or a job is served outside its legitimate service duration, the attacker can be easily detected, e.g., by the corresponding consumer. Under this formulation, the case $\beta = 1$ is of special interest to us as Problem (\ref{pr:maxmin}) can be transformed into a \emph{maximization} problem. To see this, consider any undetectable strategy followed by the attacker such that $a'_j = d'_j = t_j$,  for some $t_j \in [a_j,d_j]$, for \emph{all} jobs $j\in J$. All such strategies are always feasible to the attacker by our assumption of $\beta=1$ and, if employed by the attacker, leave no degrees of freedom to the controller. Moreover, due to the monotonicity and convexity of $C$, it is easy to see that it suffices for the attacker to consider only this set of strategies. Therefore the Problem (\ref{pr:maxmin}), under ($\beta=1$), reduces to a cost maximization problem which looks for a strategy that serves each job in a single feasible time slot. The attacker hence solves the following problem:
\vspace{-0.05in}
\begin{align}
C_{max}(a,d,e) & =  \max_{S}
 \sum_{t=1}^{T} C(E_S(t)) \notag\\
 \text{s.t.} \hspace{1em}
&  s_{jt} = 0, \hspace{2em} \forall j \in J, \forall t \in [0,T], t\neq t_j \notag\\
&  s_{jt_j} = e_j, \hspace{2em} \forall j \in J, \notag\\
&  t_j \in [a_j,d_j], \hspace{2em} \forall j \in J. \tag{Pmax}\label{pr:max}
\end{align}

We provide efficient offline and online solutions to Problems (\ref{pr:min}) and (\ref{pr:max}) in the next section, and upper and lower bounds for Problem (\ref{pr:maxmin}) in Section~\ref{sec:limited_attacks}. For comparison purposes, we will also consider an inelastic scheduling policy for the controller as a baseline, where each job is served its total energy immediately upon its arrival. This strategy represents the case when the delay tolerance of the jobs is not exploited. Therefore, the resulting cost resembles that paid in the \emph{current regular gird}, where no communication channels are established, and accordingly, the system is not vulnerable to the cyber-attacks discussed in this paper. The resulting baseline cost is defined as: $C_{base}(a,d,e) = \sum_{t\in [0,T]} C\left( \sum_{j:a_j = t} e_j \right).$

Finally, the following definitions are used throughout this paper. For each job $j\in J$, define its job allowance to be $l_j = d_j-a_j$ and let $l_{max} = \max_{j\in J} l_j$, $l_{min} = \min_{j\in J} l_j$, and $e_{max} = max_{j \in J} e_j$. Denote the set of the endpoints of the job intervals by $X \colon = \{a_1,\ldots,a_n\} \cup \{d_1,\ldots,d_n\}=\{1,\ldots,q\}$. For every pair $k \leq l \in X$, let $\mathcal{I}(k,l)$ be the set of all jobs whose intervals are entirely contained in $[k,l]$, that is, $\mathcal{I}(k,l) = \{j\in J \colon a_j \geq k, d_j \leq l\}$.

\section{Optimal Strategies and Performance Bounds}\label{sec:optimal_strategies}
In this section, we first find the optimal scheduling strategy for the controller (the solution to Problem (\ref{pr:min})). Second, we study Problem (\ref{pr:max}) and propose both an optimal offline attack and a simple online attack and compare their performance. Finally, an explicit bound on the impact of an attack is presented.

\subsection{Optimal Scheduling for the Controller}\label{sec:optimal_strategies_min}
The optimization problem at the controller (Problem \ref{pr:min}) can be directly mapped to the ``minimum-energy CPU scheduling problem" studied in \cite{Yao1995}. Our discussion below is an adapted discrete-time version to that of \cite{Yao1995}. Define the \emph{energy intensity} on $\mathcal{I}(k,l)$ to be
\vspace{-0.05in}
\begin{equation}\label{eq:intensity}
    E(\mathcal{I}(k,l)) = \frac{\sum_{j \in \mathcal{I}(k,l)} e_j}{l-k+1},
\end{equation}
\noindent and let $\mathcal{I}^*(k^*,l^*)$ be the set of jobs that maximizes $E(\mathcal{I}(k,l))$ over all $k,l \in X$. It is shown in \cite{Yao1995} that the optimal strategy schedules a total energy of $E(\mathcal{I^*}(k^*,l^*))$ in each time slot in $[k^*,l^*]$. Hence a greedy algorithm that searches for $\mathcal{I}^*$, schedules the jobs in $\mathcal{I}^*$ and then removes those jobs (and the corresponding interval) from the problem instance, can be used to solve Problem (\ref{pr:min}):
\vspace{+1pt}
\begin{alg}\label{alg:min_offline}
Repeat the steps below until $J$ is empty.
\begin{enumerate}
  \item Identify $\mathcal{I}^*(k^*,l^*)$. Schedule the jobs in $\mathcal{I}^*(k^*,l^*)$, such that $E_S(t) = E(\mathcal{I}^*(k^*,l^*))$, for all $t \in [k^*,l^*]$, according to the Earliest Deadline First (EDF) policy (which is always feasible).
  \item Modify the problem to reflect the deletion of the jobs in $\mathcal{I}^*$: For all jobs $j\in J \setminus \mathcal{I}^*$, if $a_j \geq k^*$, set $a_j \leftarrow \max(k^*-1, a_j - (l^*-k^*) -2)$; modify $d_j$ similarly. Set $J\leftarrow J \setminus \mathcal{I}^*$.
\end{enumerate}
\end{alg}
\vspace{+1pt}

The above algorithm arrives at the optimal schedule with complexity $O(n^2)$. In our simulations (Section \ref{sec:numerical}), we also compute an upper bound on the solution using an online algorithm that simply distributes the energy requirement of each job \emph{evenly} on its service interval \cite{Yao1995}.

\subsection{The Fully-compromised Controller}\label{sec:optimal_strategies_max}
We now turn our attention to Problem (\ref{pr:max}) and form a graph theoretic version of this problem. This is useful for describing the optimal full attack strategy, and for studying the impact of more limited attacks. Let $G = (V,E)$ be the interval graph induced by the jobs in $J$, i.e., each vertex $v_j \in V$ corresponds to a job interval, given by $[a_j,d_j]$, while an edge is thrown between any two vertices iff the two corresponding job intervals intersect at one or more time slots \cite{Gross2006}. We define the corresponding cost function over subsets of $V$ as $f\colon 2^V \rightarrow \mathbb{R}^+$, given by $f(\emptyset) = 0, f(S) = C \left(\sum_{j \in S} e_j\right)$ for any $S \subseteq V$. In the induced interval graph, a \emph{clique} is a subset of vertices $S \subseteq V$, such that every two vertices in $S$ are connected by an edge. A \emph{maximal clique} (inclusion-wise) is a clique that is not a subset of a larger clique. By these definitions, our problem corresponds to finding a \emph{clique partition} of $G$ that maximizes the total cost taken over the cliques in this partition, i.e., find
\vspace{-0.05in}
\begin{equation}\label{eq:max_problem}
    C_{max}(G) = \max_{\mathcal{Q} \in \mathcal{P}(G)} \sum_{K \in \mathcal{Q}} f(K),
\end{equation}
\noindent where $\mathcal{P}(G)$ is the set of all clique partitions of $G$. By our assumptions on $C$, the set function $f$ is non-decreasing, i.e., $f(S)\leq f(T),$ whenever $S \subseteq T \subseteq V$. Moreover, for every $S, T \subseteq V$ such that $f(S)\geq f(T)$, and $u \in V\setminus (T\cup S)$, we have $f(S+u) + f(T) \geq f(S) + f(T+u)$. By these two properties, the optimal clique partition, solving (\ref{eq:max_problem}), includes a maximal clique \ignore{\footnote{A similar fact is proved in \cite{Gijswijt2007}, where the authors consider clique partitioning so as to \emph{minimize} a submodular cost function on the cliques. Interestingly, the existence of a maximal clique in the optimal partition holds for both the problem considered in \cite{Gijswijt2007} and our problem.}}. In fact, if we let $G(\mathcal{I}(k,l))$ be a subgraph of $G$ restricted only to the jobs in any $\mathcal{I}(k,l)$, then $C_{max}(G(\mathcal{I}(k,l)))$ is achieved by a partition that contains a maximal clique of the subgraph $G(\mathcal{I}(k,l))$ as well. Hence, for any such subgraph, each maximal clique contained in the subgraph \emph{separates} the optimization problem into two subproblems and a Dynamic Programming algorithm (adapted from~\cite{Gijswijt2007}) solves the problem accordingly.

Let $\overline{C}(k,l)$ be the maximum feasible cost that could be achieved by scheduling the jobs in $\mathcal{I}(k,l)$. Given $k<z<l$, let $K_{k,l}^z$ be the set of all the jobs whose intervals contain time slot $z$, i.e., $K_{k,l}^z$ is a maximal clique contained in $\mathcal{I}(k,l)$. By our discussion, the following recursion clearly holds.
\vspace{-0.05in}
 \small
\begin{gather}\label{eq:max_recursion}
  \overline{C}(k,l) = \max_{z\in[k,l]} \left[ C \left( \sum_{j\in K_{k,l}^z} e_j \right)
      + \overline{C}(k,z-1) + \overline{C}(z+1,l) \right].
\end{gather}
\normalsize

Our algorithm iterates over all intervals $[k,l], k,l\in X, k\leq l$, with increasing interval length. In each iteration step, the algorithm computes $\overline{C}(k,l)$, where the last two terms are obtained from previous iterations. A formal description of this Dynamic Program is now presented.
\vspace{+1pt}
\begin{alg}\label{alg:max_offline}\cite{Gijswijt2007}
For all $k\in X$, set the initial condition
\vspace{-0.05in}
  \begin{equation}
     \overline{C}(k,k) = C \left( \sum_{j\in \mathcal{I}(k,k)} e_j \right).
  \end{equation}
With increasing subproblem width $(l-k)$, apply the following Dynamic Program:
\begin{enumerate}
  \item Solve the optimization (\ref{eq:max_recursion}). Denote the solution by $z^*$.
  \item Update the clique partition
  \vspace{-0.05in}
  \begin{equation*}\label{eq:rec02}
    \mathcal{Q}(k,l) = \begin{cases}
    \emptyset, \mbox{  if } \mathcal{I}(k,l) = \emptyset,\\
    \mathcal{Q}(k,z^*-1) \cup K_{k,l}^{z^*} \cup \mathcal{Q}(z^*+1,l), \mbox{  o.w.}
    \end{cases}
  \end{equation*}
\end{enumerate}
\end{alg}
\vspace{+1pt}

The optimal cost is $C(1,q)$ and the optimal clique partition is $\mathcal{Q}(1,q)$, which are computed in the final step of the above program. From the obtained clique partition, one can easily compute a set of time slots, $t_j, j\in J$ and set $a'_j = d'_j = t_j$, solving Problem (\ref{pr:max}). The obtained schedule leaves no degrees of freedom to the controller as, after the attacker's modifications, all jobs become virtually urgent to controller and must be scheduled immediately upon their arrival. It is also clear that, as the jobs' allowance increases, the attacker is capable of forming larger cliques and hence imposing higher costs on the controller. Our goal in the remainder of this section is to formalize this observation. Towards this end, we first present a simple online attack where the jobs in $J$ are partitioned into cliques according to an EDF policy. That is, starting from the earliest deadline, all the jobs that arrive before or upon each deadline are grouped in a single clique and then removed from the problem instance:
\vspace{+1pt}
\begin{alg}\label{alg:max_online}

Set $i=1,m=0$. Repeat until $J$ is empty:
\begin{enumerate}
  \item Set $\tilde{d} = \min_{j \in J} d_j$.
  \item Set $N_i = \{j\in J \colon a_j \leq \tilde{d} \}$.
  \item For all $j\in N_i$, set $t_j = \tilde{d}$.
  \item Update $J \leftarrow J\setminus N_i$, $m\leftarrow m+1, i\leftarrow i+1$.
\end{enumerate}
\end{alg}
\vspace{+1pt}

Once the clique partition $\{N_i\}, i\in\{1,\ldots,m\}$, has been established, the resulting cost is computed as
\vspace{-0.05in}
\begin{equation}\label{eq:max_bound}
    \underline{C}_{max} = \sum_{i=1}^m  C \left(\sum_{j \in N_i} e_j\right).
\end{equation}

Our next result shows that, despite its simplicity and online operation, Algorithm \ref{alg:max_online} could still achieve a significant loss in the system's efficiency:
\vspace{+1pt}
\begin{prop}\label{prop:max_online_tightness}
For $C(E)=E^b, b\in \mathbb{R}, b\geq 1$, Algorithm \ref{alg:max_online} has an approximation factor of $\frac{1}{r^{b-1}}, r \colon = \lceil \frac{l_{max}}{l_{min}}\rceil + 1$.
\end{prop}
\vspace{+1pt}
Moreover, when $C(.)$ is a power function of the form $C(E)=E^b, b\in \mathbb{R}, b\geq 1$, the simple structure of the online solution also allows us to arrive at an explicit lower bound for $C_{max}$:
\vspace{+1pt}
\begin{prop}\label{prop:max_lower_bound}
For $C(E)=E^b, b\in \mathbb{R}, b\geq 1$,
\vspace{-0.05in}
\begin{equation}
    C_{max}(a,d,e) \geq \left( \frac{l_{min}\sum_{j \in J} e_j}{2 l_{min} + a_n - a_1} \right)^b.
\end{equation}
\end{prop}
\vspace{+1pt}

\begin{figure}
 \centering
 \includegraphics[width=0.45\textwidth]{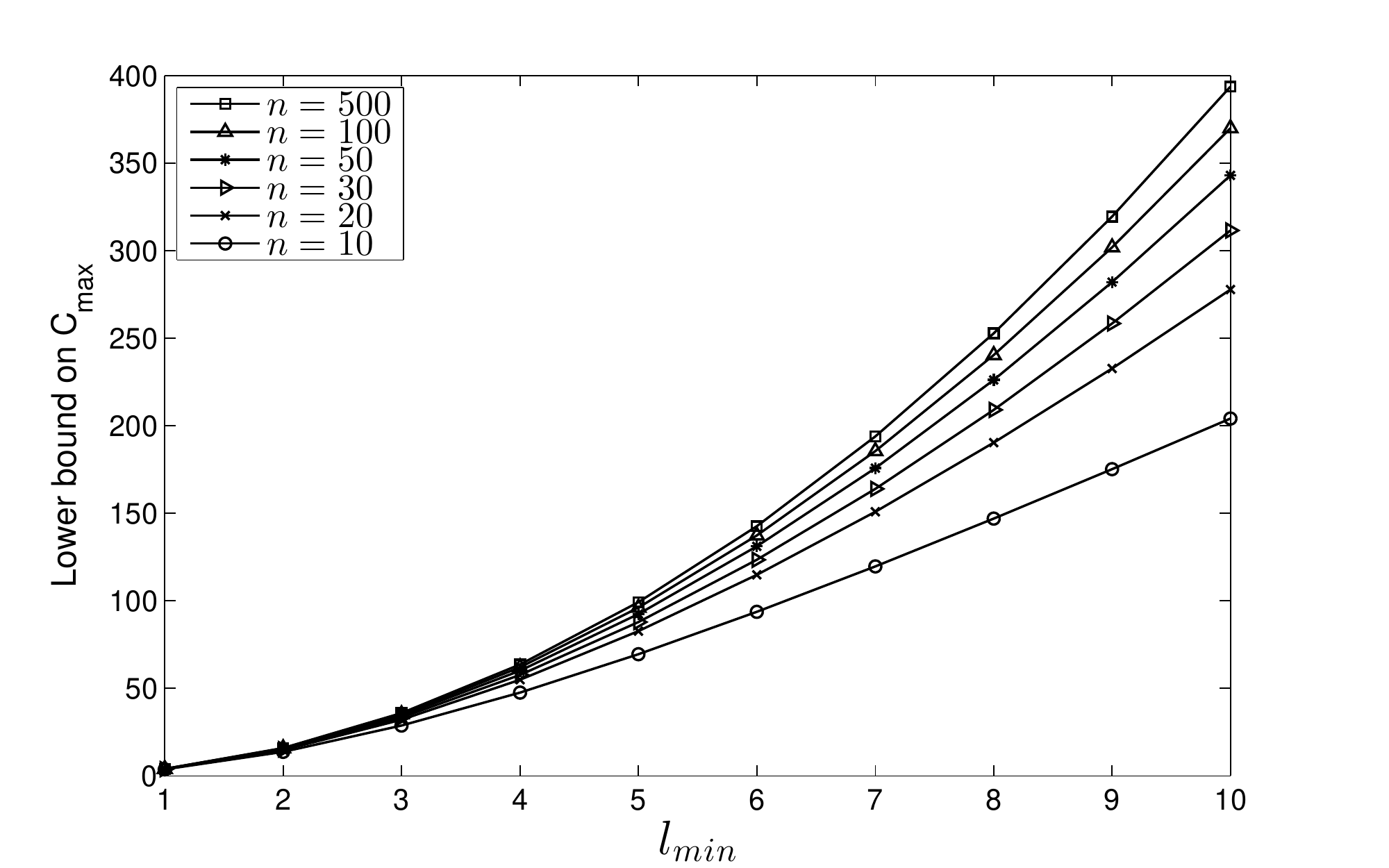}
 \caption{A lower bound on $C_{max}$ plotted for various values of $n$ and $l_{min}$ under a quadratic cost function (i.e., $b = 2$). The average energy demand is 10 while the average inter-arrival time is 5.}
 \label{fig:max_lower_bound}
\end{figure}

The proofs for both propositions are provided in the Appendix. The above result can be used to estimate the growth of $C_{max}$ with $l_{min}$. For instance, if we fix the average energy demand and the average inter-arrival time to arbitrary values, the bound obtained in Proposition \ref{prop:max_lower_bound} versus an increasing $l_{min}$ can be plotted. See Figure \ref{fig:max_lower_bound} for an example. As shown in the figure, $C_{max}$ grows at least linearly with $l_{min}$, and the rate of growth increases as the sample size $n$ increases. More numerical results are reported in Section \ref{sec:numerical}.

Our numerical results in Section \ref{sec:numerical} provide more insights on the performance of the online attack.

\section{Performance Bounds under Limited Attacks}\label{sec:limited_attacks}
In this section, we study the case where the attacker is capable of changing the arrival times and the deadlines of only $B = \beta n$ jobs. Similar to our argument in Section \ref{sec:formulation}, the attacker could only consider the following strategy: Choose a set of jobs $J^* \subset J$ such that $|J^*|=\beta n$, and set $a'_j = d'_j = t^*_j$ for all jobs $j\in J^*$ and leave all other jobs unaltered. We propose two polynomial time algorithms that render a lower and an upper bound, respectively, on the performance due to the considered limited attack. For simplicity, we let $C_{max} = C_{max}(a,d,e)$ and $C_{maxmin}(\beta) = C_{maxmin}(a,d,e,\beta)$.

\subsection{A lower bound}
Inspired by the standard greedy algorithm for the fractional knapsack problem~\cite{CLRS}, we propose a simple variant that is tailored to our problem. In the classical fractional knapsack problem, $m$ items are given, each with a weight $w_i$ and a value $v_i$. We need to specify which items to collect such that their total weight does not exceed a specified quantity ($\beta_0 \sum_{i} w_i, 0\leq \beta_0 \leq 1$) and their total value is maximized. A fraction of any item might be collected, and the corresponding value is scaled according its chosen weight. The greedy algorithm below solves this problem.
\vspace{+1pt}
\begin{alg}\label{alg:greedy}
Given $(v_1,w_1),\ldots,(v_m,w_m)$ and $\beta_0$
\begin{enumerate}
  \item Sort $(v_i,w_i)$ according to $v_i/w_i$ in a non-increasing order.
  \item Choose the first $k$ pairs, $(v_1,w_1),\ldots,(v_k,w_k)$ s.t.
  \vspace{-0.05in}
  \begin{equation}
      \sum_{i=1}^k w_i \leq \beta_0 \sum_{i=1}^m w_i, \hspace{3em}  \sum_{i=1}^{k+1} w_i > \beta_0 \sum_{i=1}^m w_i.
  \end{equation}
\end{enumerate}
\end{alg}
\vspace{+1pt}

The optimal set is the chosen $k$ items in step (2), and a fraction of the $k+1$-th item as the budget allows. Moreover, if we let the remaining weight budget after selecting the first $k$ pairs to be $\beta_1$ $= \left(\beta_0 \sum_{i=1}^m w_i - \sum_{i=1}^k w_i\right) / w_{k+1}$, by the greedy selection, we have
\vspace{-0.05in}
\begin{equation}\label{eq:fractional-knapsack}
    \sum_{i=1}^k v_i + \beta_1 v_{k+1} \geq \beta_0 \sum_{i=1}^m v_i.
\end{equation}

The proposed attack strategy builds on this algorithm: first, the attacker finds the optimal clique partition using Algorithm \ref{alg:max_offline}, assuming a full budget. Then, it utilizes the above algorithm twice; once to choose a set of cliques to fully compress (i.e., to collapse the job allowances within each clique to one common time slot), and to choose a set of jobs within a given clique to fully compress. The choice that results in a higher cost is adapted.
\vspace{+1pt}
\begin{alg}\label{alg:maxmin_greedy}
\begin{enumerate}
  \item Find the optimal clique partition of the jobs, $K_1,\ldots,K_m, 1 \leq m \leq n$, using Algorithm \ref{alg:max_offline} (assuming a full budget). For each clique $K_i$, set $E_i = \sum_{j \in K_i} e_j$ and $N_i = |K_i|$.
  \item Apply Algorithm \ref{alg:greedy} to the pairs $(C(E_i),N_i), 1 \leq i \leq m$, and $\beta$, and pick the resulting $k$ cliques (ignoring the fraction generated by the algorithm). Compute the cost $C_1$ resulting from fully compressing those $k$ cliques. That is, $C_1 = \sum^k_{i=1} C(E_i)$.
  \item For the $(k+1)$-th clique, apply Algorithm \ref{alg:greedy} to the pairs $(e_j,1), j\in K_{k+1}$ and $\beta_2 \colon = \frac{\beta n}{N_{k+1}}$. Compute the cost $C_2$ resulting from fully compressing the chosen $k'$ jobs. That is, $C_2 = C \left(\sum^{k'}_{j=1} e_j\right)$.
  \item If $C_1\geq C_2$, fully compress the jobs in cliques $1,\ldots,k$. Otherwise, fully compress the chosen jobs from clique $k+1$. Set $\underline{C}_{maxmin}(\beta) = \max(C_1,C_2)$.
\end{enumerate}
\end{alg}
\vspace{+1pt}
To get insights on the performance of this attack, suppose that, under no budget constraints, the optimal clique partition (obtained from Algorithm \ref{alg:max_offline}) is composed of cliques of size one. In this case, our greedy attack will choose to fully compress the $\beta n$ jobs of the highest energy demands. This guarantees that $\underline{C}_{maxmin}(\beta) \geq \beta C_{max}$. Another extreme case is when the optimal clique partition is composed of one clique containing the $n$ jobs. Here, our greedy selection guarantees that $\underline{C}_{maxmin}(\beta) \geq C\left(\beta \sum_{j\in J} e_j\right)$. When $C(.)$ is a power function of the form $C(E)=E^b, b\in \mathbb{R}, b\geq 1$, we then get $\underline{C}_{maxmin}(\beta) \geq \beta^b C_{max}$. For more general clique partitions, these two insights are used to arrive at the achievable bound below (the proof is found in the Appendix).
\vspace{+1pt}
\begin{prop}\label{prop:maxmin_lower_bound}
For $C(E)=E^b, b\in \mathbb{R}, b\geq 1$,
\vspace{-0.05in}
\begin{equation}\label{eq:maxmin_bound}
    \underline{C}_{maxmin}(\beta) \geq \frac{\beta^b}{2} C_{max}.
\end{equation}
\end{prop}
\vspace{+1pt}

\subsection{An upper bound}
In order to compute an upper bound on the system's performance, we find the optimal attack strategy under the assumption that the controller follows {\it the baseline scheduling strategy given in Section \ref{sec:formulation}}. We further assume that at most one job arrives at any given time slot $t\in [0,T]$. Our main observation is that, under these assumptions, Problem \ref{pr:maxmin} can be solved by a Dynamic Programming algorithm similar to Algorithm \ref{alg:max_offline}. To illustrate, consider the solution to Problem \ref{pr:maxmin} when the controller follows the baseline scheduling strategy. The jobs' schedule under this solution is a clique partition of the induced graph $G$, which we denote by $\mathcal{P}^*(G)$. Let $O$ denote the set of all the cliques of size one in $\mathcal{P}^*(G)$ and $O' = \mathcal{P}^*(G) \setminus O$. Since at most one job can arrive at any time slot, without of loss of optimality, we can assume that each clique $K \in O'$ contains exactly a single job, say $j_K$, that has an unaltered arrival time. The remainder of the jobs would have arrival times altered to match that of $j_K$. For instance, we can choose $j_K$ as the job with latest arrival in clique $K$. Hence, the budget used to form clique $K$ is exactly $|K|-1$. This observation leads to the below proposition (the proof is found in the Appendix).
\vspace{+1pt}
\begin{prop}\label{prop:maxmin_DP}
Let $K_1 \in O'$ be the clique containing the maximum total energy requirement in $O'$. If $K_1$ is not a maximal clique in $G$, it can be made maximal by adding job(s) only from $O$.
\end{prop}
\vspace{+1pt}
The above proposition can be directly applied to any subgraph $G(\mathcal{I}(k,l))$, as defined in Section \ref{sec:optimal_strategies}. Similar to the case $\beta = 1$, for any such subgraph, each maximal clique contained in the subgraph separates the optimization problem into two subproblems. Hence, if we dedicate a budget of $m$ jobs to any interval $[k,l]$, we can construct a recursion that computes $\overline{C}(k,l,m)$ by parsing for maximal cliques in each time slot $z \in [k,l]$, investigating all the possibilities of using only a budget of $i$ out of $m$ for each found clique. We would also exhaust all the possibilities of distributing the remaining budget $m-i+1$ on the resulting two subproblems of any chosen clique, and any chosen budget for that clique. By Proposition~\ref{prop:maxmin_DP}, the constructed recursion indeed holds. A Dynamic Program similar to Algorithm \ref{alg:max_offline} is built and the results are reported in Section \ref{sec:numerical}.

\section{Numerical Results}\label{sec:numerical}

\begin{figure}
 \centering
 \includegraphics[width=0.45\textwidth]{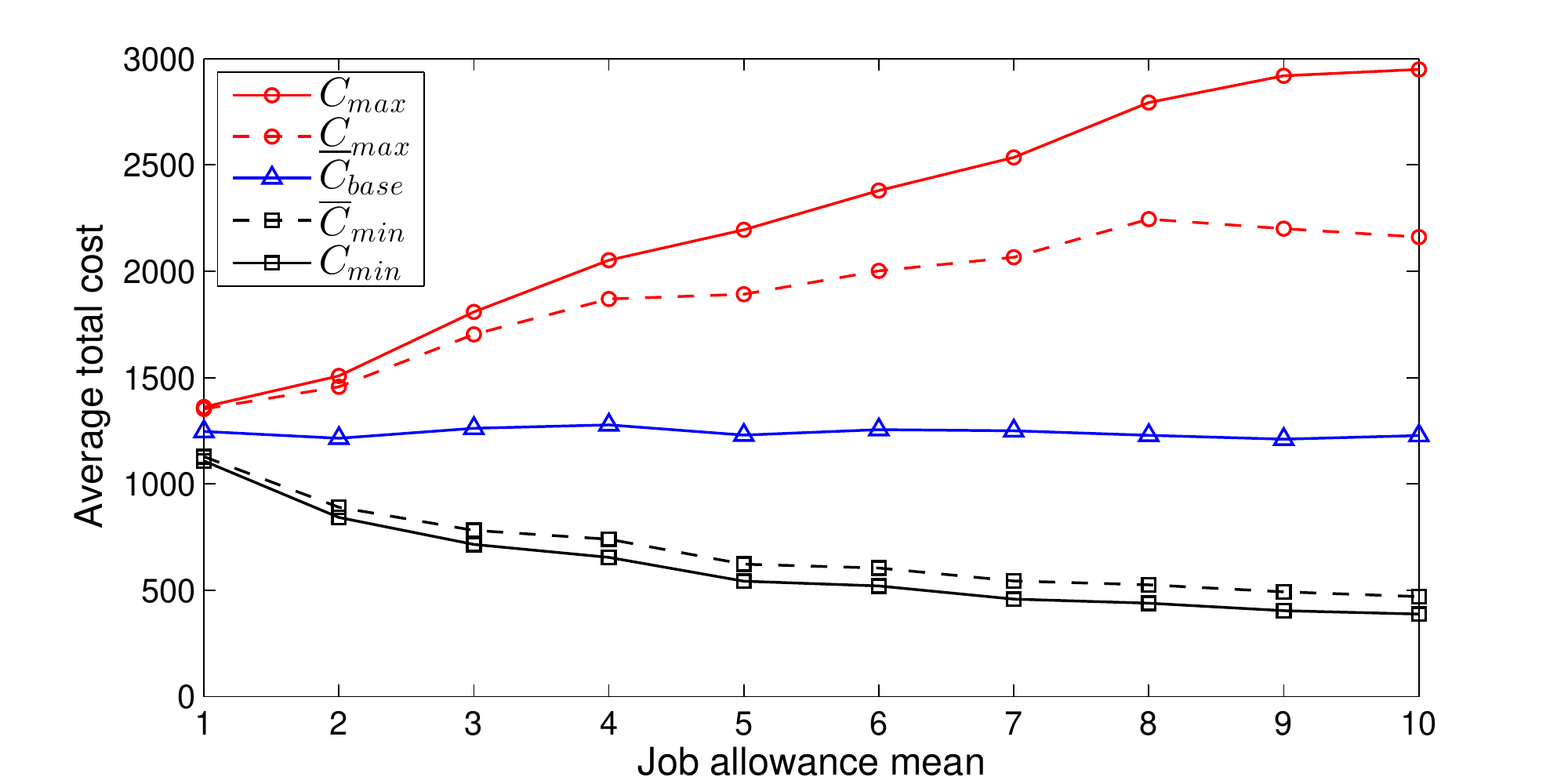}
 \caption{Average total cost for different scheduling strategies versus the job allowance mean. The mean interarrival time is 5.}
 \label{fig:bounds_var_powers}
\end{figure}

\begin{figure}
 \centering
 \includegraphics[width=0.45\textwidth]{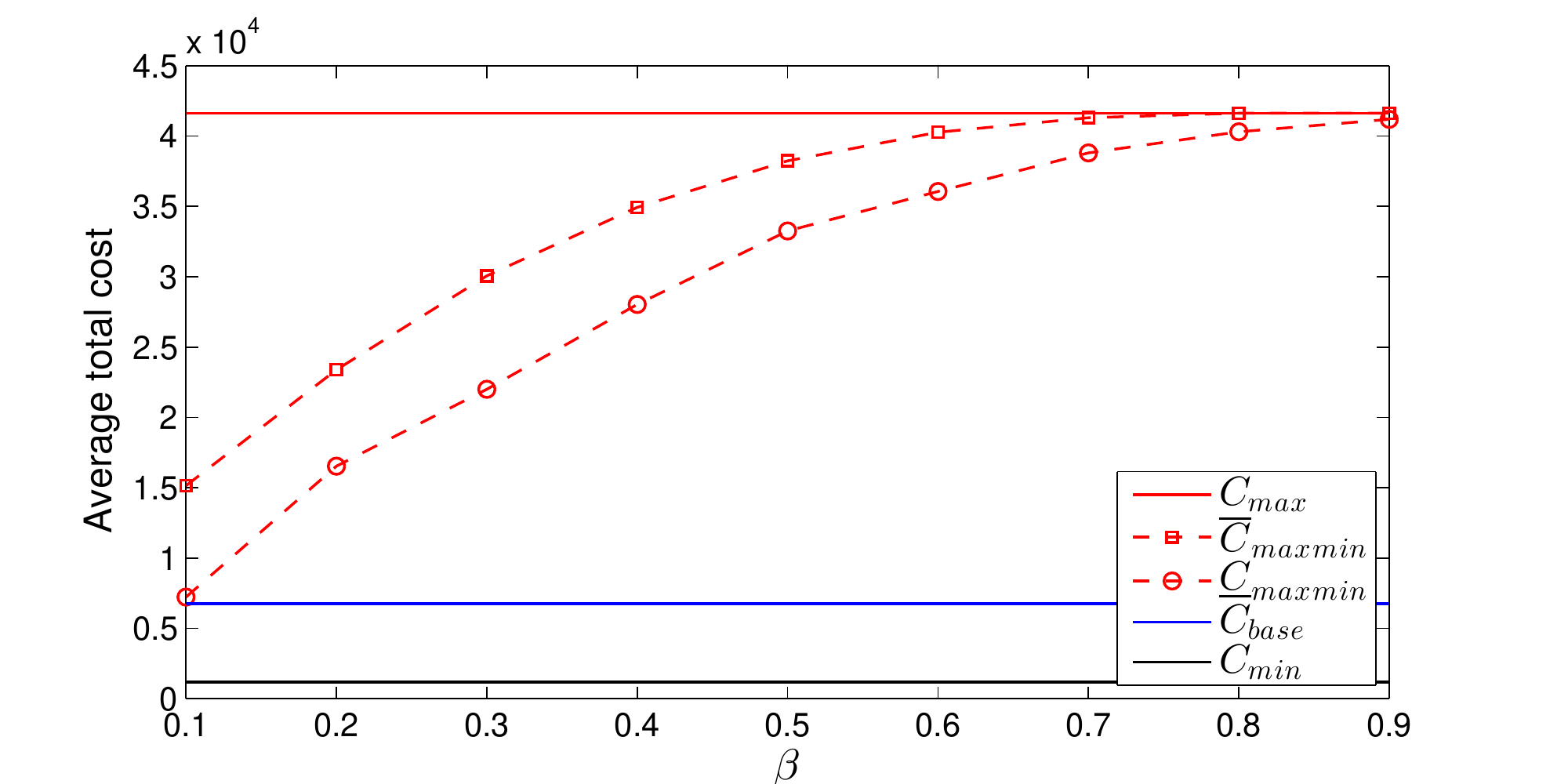}
 \caption{Bounds on the total cost attained by a partial attack with a varying $\beta$ for 50 jobs.}
 \label{fig:maxmin_bounds}
\end{figure}

In this section, the job arrivals are simulated as a Poisson arrival process with mean 5. All the job allowances are independently and identically distributed exponential random variables. We use a quadratic cost function $C(E) = E^2$ in all of our simulations. Figure \ref{fig:bounds_var_powers} reports our comparison between the maximum and the minimum cost caused by an optimal/online full attack and an optimal/online uncompromised controller, for $n=100$. The results are obtained by varying the job allowance mean and are averaged over 20 trials. The amount of energy demands is uniformly distributed on $[1,5]$. As shown, as the job allowance mean increases, more flexibility is offered to the uncompromized controller, hence enabling further cost reductions. This, however, offers a similar opportunity for the attacker to form larger cliques of jobs and increase the harm. For instance, a fully unprotected controller, on the average, ends up paying $250\%$ of the expected baseline cost ($700\%$ of the expected minimum cost), under an optimal full attack and large-enough time-flexibility. Our proposed suboptimal algorithm for the attacker maintains significant gains over both the baseline and minimum costs, even with the increased job allowance variance. Figure \ref{fig:maxmin_bounds} focuses on the performance of partial attacks that were launched using the proposed algorithms in Section \ref{sec:limited_attacks}. In our experiment, the simulation sample is composed of 50 jobs. The energy requirements were uniformly distributed on $[1,20]$ while the mean job allowance was set to 40. The results are averaged over 5 trials. As shown, with the increased allowance mean, the obtained clique partitions become denser and therefore the bounds become tighter. Also, observe that using a simple greedy algorithm, the attacker is immediately capable of achieving a cost arbitrarily close to $C_{base}$ for our sample, with a chance of altering only 5 jobs out of 50.

\begin{figure}
 \centering
 \includegraphics[width=0.45\textwidth]{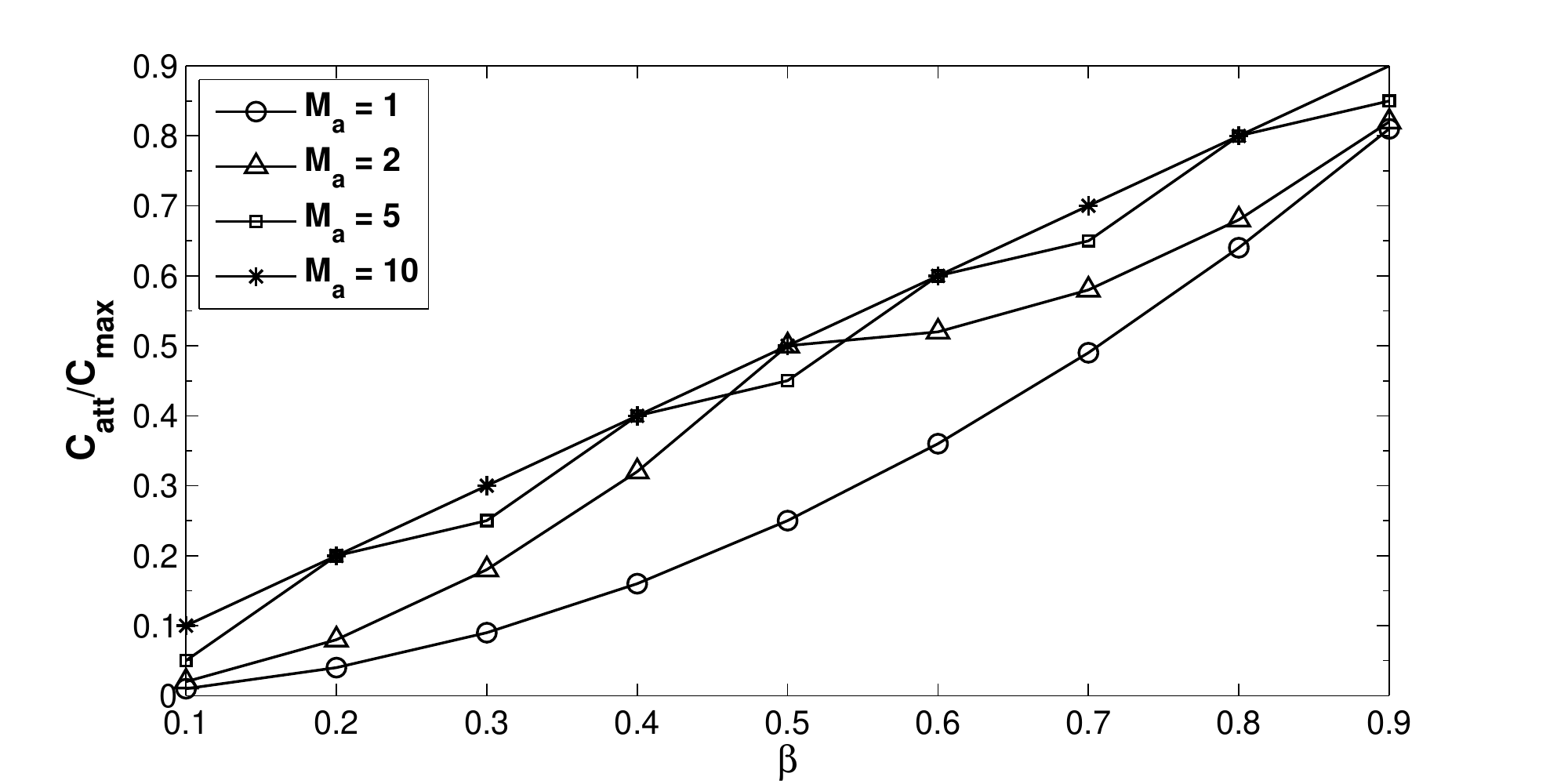}
 \caption{Performance of a limited attack with respect to the allowed budget fraction $\beta$ for 50 identical jobs with $e_j = 5, l_j = 50$. The interarrival times are all set to $M_a$. }
 \label{fig:maxmin_ordered_powers}
\end{figure}

Finally, in a more controlled experiment, we have generated 50 identical demands, with each requiring a 5 energy units and offering an allowance of 50. The interarrival times between jobs are all set to one value, denoted by $M_a$ in Figure \ref{fig:maxmin_ordered_powers}. $M_a$ was set to various values between 1 and 10, and $\underline{C}_{maxmin}/C_{max}$ was computed with varying values of $\beta$. This enables us to gain more insights on how the growth of $C_{maxmin}$, with respect to $\beta$, and how this growth is affected by the clique densities. As shown in the figure, when $M_a=1$, with our chosen parameters, a single clique of jobs could be formed to achieve the maximum cost, and hence, in accordance with our theoretical results, the attacker could achieve approximately $\beta^2$ of the maximum achievable cost. As $M_a$ increases, the growth of $\underline{C}_{maxmin}/C_{max}$ with $\beta$ approaches a linear trend. The reason is that as $M_a$ increases, the size of the optimal clique partition of jobs increases, having approximately equally sized cliques. Hence the maximum cost decreases so does the contribution of each clique to the maximum cost.

\section{Conclusion}\label{sec:conclusion}

In this paper, we have studied the performance of the smart grid, in terms of energy efficiency, in the presence of active attacks on the system. When the grid operator is fully compromised, we have proposed optimal scheduling and undetectable attack strategies. We have derived bounds on both the minimum and maximum achievable cost by an attacker with low complexity, online algorithms. In addition, we gave bounds on the impact of attacks that are limited by intrusion detection at the operator. In these limited attacks, we have shown that a significant increase in cost could still be achieved by a simple greedy algorithm. Overall, our theoretical analysis and numerical results show that an inelastic utilization of the communication channels in the smart grid could result in costs significantly higher than those expected for both the smart grid and the current electric grid, motivating the need for stronger intrusion detection and defense strategies for grid operators.

\bibliographystyle{IEEEtran}
\bibliography{refs}

\appendix
\section{Proofs}
\subsection{Proof of Proposition \ref{prop:max_online_tightness}}

For a given problem instance, $J$, $a,d,e$, let the optimal partition of the jobs in $J$ be $K_1, K_2, \ldots, K_{m^*}$ such that
\vspace{-0.05in}
\begin{equation}
    C_{max}(a,d,e) = \sum_{z=1}^{m^*} \left( \sum_{j\in K_z} e_j \right)^b.
\end{equation}
By our construction of $\{N_i\}, i\in\{1,\ldots,m\}$, we have that, for all $i'>i$, all the jobs in $N_{i'}$ have arrived strictly later than the earliest deadline of the jobs in $N_i$. Consequently, each $K_z, z\in \{1,\ldots,m^*\}$ could have a nonempty intersection with at most $r$ \emph{consecutive} sets in the partition $\{N_i\}, i\in\{1,\ldots,m\}$. Letting $K(z,i) = K_z \cap N_i$, we have
\vspace{-0.05in}
\begin{eqnarray*}
    C_{max}(a,d,e)  &=& \sum_{z=1}^{m^*} \left( \sum_{i=1}^m \left( \sum_{j\in K(z,i)} e_j \right) \right)^b\\
                    &\overset{(a)}{\leq}& \sum_{z=1}^{m^*} r^{b-1} \sum_{i=1}^m  \left( \sum_{j\in K(z,i)} e_j \right)^b\\
                    &\leq& r^{b-1} \underline{C}_{max}(a,d,e),
\end{eqnarray*}
\noindent where (a) is obtained by the power mean inequality.

\subsection{Proof of Proposition \ref{prop:max_lower_bound}}

From Eq(\ref{eq:max_bound}) and the power mean inequality, we have
\vspace{-0.05in}
\begin{equation}
    C_{max} \geq \underline{C}_{max} \geq \left(\frac {\sum_{j\in J} e_j}{m}\right)^b.
\end{equation}
We will show that $m \leq n/r +2$, where $r = \frac{n l_{min}}{a_n-a_1}$, and this completes the proof. If $r\leq 1$, we are done. Otherwise, it suffices to show that $m \leq n/r +2$ for $(n/r) \in \mathbb{R}\setminus \mathbb{N}^+$.

In the solution of Algorithm \ref{alg:max_online}, the maximum number of cliques of size 1 is $\lfloor n/r\rfloor$, for otherwise, our assumption on $r$ is violated. We assume that the number of cliques of size 1 is $\lfloor n/r\rfloor - k, k \geq 0$. Hence the summation of the interarrival times of those jobs is GE $\left( \lfloor n/r\rfloor - k \right) l_{min}$, if they did not include the last arrival in $J$, and is GE  $\left( \lfloor n/r\rfloor - k -1 \right) l_{min}$ if they did. On the other hand, if the number of the remaining cliques is strictly larger than $k+2$, then necessarily the summation of the interarrival times corresponding to those cliques is strictly larger than $(k+1) l_{min}$, if they include the last arrival, and strictly larger than $(k+1) l_{min}$ otherwise. Combined with the argument above, we find that we can have at most $k+2$ remaining cliques, and accordingly $m \leq \lfloor n/r \rfloor +2 \leq n/r +2$.

\subsection{Proof of Proposition \ref{prop:maxmin_lower_bound}}

Let $\beta_1 = \frac{\beta n - (N_1+...+N_k)}{N_{k+1}}$ denote the fraction of budget available to clique $k+1$ assuming the first $k$ cliques are fully compressed, and $\beta_2 = \beta \frac{n}{N_{k+1}}$.

By the greedy selection of jobs in step (3) of Algorithm~\ref{alg:maxmin_greedy} and~\eqref{eq:fractional-knapsack}, we have $\sum^{k'}_{j=1} e_j \geq \beta_2 E_{k+1}$. Therefore, $C_2 \geq \beta_2^b E^b_{k+1}.$ Let $C_0 = C_1 + \beta_1 E^b_{k+1}$. Then by the greedy selection of cliques in step (2) of Algorithm~\ref{alg:maxmin_greedy} and~\eqref{eq:fractional-knapsack}, we have $C_0 \geq \beta \sum_{i=1}^m E^b_i = \beta C_{max}.$ We then have
\vspace{-0.05in}
\begin{eqnarray*}
\frac{\underline{C}_{maxmin}}{C_0} &=& \frac{\max(C_1,C_2)}{C_1 + \beta_1 E^b_{k+1}}
\geq \frac{C_2}{C_2 + \beta_1 E^b_{k+1}} \\
&=& \frac{\beta_2^b E^b_{k+1}}{\beta_2^b E^b_{k+1} + \beta_1 E^b_{k+1}}
= \frac{\beta^b_2}{\beta^b_2+\beta_1}  \\
&\overset{(a)}{\geq}& \frac{\beta^b_2}{\beta^b_2+\beta_2}
= \frac{\beta_2^{b-1}}{\beta_2^{b-1}+1}  \\
&\overset{(b)}{\geq}& \frac{\beta^{b-1}}{\beta^{b-1}+1}
\geq \frac{\beta^{b-1}}{2} ,
\end{eqnarray*}
\noindent where (a) follows from $\beta_1 \leq \beta_2$ and (b) follows from $\beta_2 \geq \beta$. Hence $\underline{C}_{maxmin} \geq \frac{\beta^{b-1}}{2} C_0 \geq \frac{\beta^b}{2} C_{max}$.

\subsection{Proof of Proposition \ref{prop:maxmin_DP}}

Assume that $K_1$ is not maximal. Then there exists a job $j \in V \setminus K_1$ such that $K_1+j$ is a clique in $G$. If $j$ is in some clique $K_2 \in O'$ and $a'_j \neq a_j$ then we can schedule job $j$ in $K_1$ without affecting the attacker's budget. Moreover, by the convexity of $C(.)$, the resulting cost cannot decrease by this change. If $j$ is in some clique $K_2 \in O'$, and $a'_j = a_j$, then we can schedule all the jobs in clique $K_1$ together with job $j$, and schedule all the jobs in $K_2$ at the latest arrival time of the remaining jobs in $K_2$. This leaves the budget unaffected and could only increase the total resulting cost.

\end{document}